\documentclass{elsart}
\usepackage{graphicx,amssymb}
\usepackage{epsfig}
\usepackage{color}
\begin{document}
\begin{frontmatter}

\title{Search for solar hadronic axions produced by a bremsstrahlung-like
       process}

%\author{D. Kekez}
%\author{A. Ljubi\v{c}i\'{c}}
%\author{Z. Kre\v{c}ak} 
%\author{M. Kr\v{c}mar}

%\affiliation{Rudjer Bo\v{s}kovi\'{c} Institute, P.O.Box 180, 10002 Zagreb,
%             Croatia}
%%%%%%%%%%%%%%%%%%%%%%%%%%%%%%%%%%%%%%%%%%%%%%%%%%%%%%%%%%%%%%%%%%%%%%%%%%
\author{D. Kekez},
\author{A. Ljubi\v{c}i\'{c}},
\author{Z. Kre\v{c}ak}, 
\author{M. Kr\v{c}mar\corauthref{cor}}
\corauth[cor]{Corresponding author.}
\ead{Milica.Krcmar@irb.hr}

\address{Rudjer Bo\v{s}kovi\'{c} Institute, P.O.Box 180, 10002 Zagreb,
             Croatia}
%%%%%%%%%%%%%%%%%%%%%%%%%%%%%%%%%%%%%%%%%%%%%%%%%%%%%%%%%%%%%%%%%%%%%%%%%% 

%\date{\today}

\begin{abstract}
We have searched for hadronic axions which may be produced in the Sun by
a bremsstrahlung-like process, and observed in the HPGe detector by an
axioelectric effect. 
A conservative upper limit on the hadronic axion mass of
$m_a \lesssim 334$~eV at 95\% C.L. is obtained.   
Our experimental approach is based on the axion-electron coupling and it does
not include the axion-nucleon coupling, which suffers from the large 
uncertainties related to the estimation of the flavor-singlet axial-vector
matrix element.   
 
\end{abstract}

\end{frontmatter}

%\pacs{14.80.Mz}

%\maketitle

The axion, a light pseudoscalar particle associated with the spontaneous
breaking of a $U$(1) Peccei-Quinn symmetry \cite{Pecc77}, was introduced 
to explain the absence of CP violation in strong interactions. The mass of the
axion satisfies  $m_a = 6\, {\rm eV}\times 10^6 \,{\rm GeV}/f_a$ \cite{Yao08}.
The symmetry-breaking scale (or the axion decay constant) $f_a$, however, is
left undetermined in the theory. The present astrophysical and cosmological
considerations \cite{Yao08} have placed bounds on the parameter, and are
consistent with $10^{-5}\, {\rm eV} \lesssim m_a \lesssim 10^{-2}\, {\rm eV}$. 
At the lower
end of this constraint, axions are a viable cold dark matter candidate, and 
experimental attempts to detect their presence are in progress \cite{Duff05}.
Besides this range of allowed axion masses, for hadronic 
axions\footnote{Although hadronic axions
do not couple directly to ordinary quarks and charged leptons, their coupling 
to nucleons and electrons is not zero due to the axion-pion mixing and 
radiatively induced coupling to electrons. Their coupling to photons is model 
dependent. It was shown by Kaplan \cite{Kap85} that is possible to construct 
hadronic axion models in which the axion-photon coupling is significantly 
reduced and may actually vanish.} 
\cite{Kim79}
there exists another window of
$10\, {\rm eV} \lesssim m_a \leq 20\, {\rm eV}$ \cite{Yao08}, as long as the
axion-photon coupling is sufficiently small \cite{Kap85}.
These bounds arising from arguments concerning
the supernova 1987A cooling and axion burst, however, are model dependent and
with large uncertainties. In respect of the early universe, axions in this
hadronic axion window can reach thermal equilibrium before the QCD phase
transition and hence, like neutrinos, they are also candidates for hot dark
matter \cite{Mor98,Hann07}.

As axions couple to photons, electrons and nucleons, the Sun would be a
strong axion emitter. Hadronic axions of continuous energy spectrum, with
an average energy of 4.2 keV, could be produced abundantly in the solar core
by the Primakoff conversion of thermal photons in the electric
fields of charged particles in the plasma. The ongoing CAST experiment at 
CERN searches directly for these axions by pointing a 
decommissioned Large Hadron Collider prototype magnet (with a field of 9 T and
length of 9.26 m) toward the Sun. Thanks to the powerful magnet and the 
installed x-ray focusing mirrors (which reduce noise), the detection 
sensitivity is considerably improved relative to previous experiments of this
kind \cite{Laz92}. The obtained upper limit
on the axion-photon coupling of $8.8\times10^{-11}$\, GeV$^{-1}$ 
\cite{And07} also
supersedes, for a broad range of axion masses, the previous limit derived from 
energy-loss arguments on globular cluster stars \cite{Yao08}.  
In the case of hadronic axions with strongly suppressed photon 
couplings the Primakoff rate is negligible.
Some nuclear processes have been proposed as sources of solar 
monoenergetic axions \cite{Raff82,Hax91} and experiments based on the 
detection of hadronic axions with
suppressed axion-photon couplings were reported by several authors
\cite{Krc98,Krc01,Lju04,Jak04,Der05,Nam07,Der07,Bel08}. 
To date the best experimental limit on 
 mass of these axions is set to be around 216 eV \cite{Nam07}. However,
as pointed by the author himself, this experimental approach based on the
axion-nucleon coupling suffers from a poorly constrained flavor-singlet 
axial-vector matrix element that
affects the axion-nucleon interaction strongly. For example,
if a recent value of this matrix element of $\simeq 0.3$ \cite{Lea05} is 
taken into account, the 
obtained result becomes weaker, $m_a < 515$~eV (95\% C.L.) \cite{Nam07}. 

\begin{figure}[htb!]
\centerline{\includegraphics[width=90mm,angle=0]{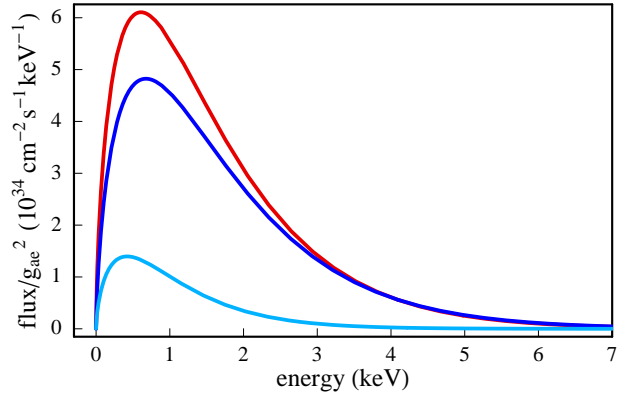}}
               \caption{
Differential solar axion flux at the Earth, derived by integrating  
Eq.~(\ref{brem}) over SSM~\cite{Bach05} up to $r=R_{\odot}$ (red line),
$r=0.2\,R_{\odot}$ (blue line), and from $r=0.2\,R_{\odot}$ to $r=R_{\odot}$
(light blue line). The axion-electron coupling $g_{ae}$
is defined in Eq.~(\ref{coupling}).}
      \label{flux}
   \end{figure}

Here we focus our attention to emission of
the hadronic axions via their radiatively induced coupling to electrons 
\cite{Sred85} in
the hot solar plasma and only the bremsstrahlung-like process, 
$e^- + Ze(e^-) \rightarrow e^- + Ze(e^-) + a$,
%$e^- + Ze \rightarrow e^- + Ze + a$, 
is used as a source of axions. In addition, 
following the calculations in \cite{Raff86}, 
we have estimated that the 
contributions of electron-electron collisions to the axion bremsstrahlung 
emission from the Sun are negligible in the 2.0~-~3.8 keV energy region 
which is of interest in our experiment. Therefore, only scattering
of electrons on protons and He nuclei was considered.
Using theoretical predictions of Zhitnitsky and Skovpen \cite{Zhi79} for
the axion bremsstrahlung due to electron-nucleus   
collisions, for the case where the Born approximation is valid 
and 
$E\gg m_a$ ($E$ is the total energy of axion), 
we can write the differential solar axion flux at the Earth as
%%%%%%%%%%%%%%%%
\begin{equation}
   \frac{d\Phi_a}{dE}= \frac{1}{4\pi d_{\odot}^2}
                         \int_0^{R_\odot}4\pi r^2 dr
        \int_0^{\infty}(N_{\rm H}+ 4N_{\rm He})\,n_e \,v_e \,
        \frac{d\sigma_a}{dE} \,dT_e\:. 
      \label{brem}
\end{equation}
%%%%%%%%%%%%%%%%
Here $d_{\odot}$ is the average distance between the
Sun and the Earth, $R_\odot$ denotes the solar radius while $N_{\rm H}$ and
$N_{\rm He}$ are the number density of hydrogen and helium nuclei 
in a given spherical shell in the solar interior
at the radius $r$, respectively. The Maxwellian distribution at the temperature
$T(r)$ for the nondegenerate, nonrelativistic incident electrons of velocities
$v_e$ and kinetic energies $T_e$ is given by    
\begin{equation}
  n_e = \frac{2\sqrt{T_e}\,e^{-T_e/kT}}
        {\sqrt{\pi}\,{(kT)}^{3/2}}\,N_e\:,
\label{elden}
\end{equation} 
where $N_e$ is the number density of the electrons at the radius $r$,
and $k$ is the Boltzmann constant. The differential 
cross section for the axion bremsstrahlung process (for $Z$=1) \cite{Zhi79} 
is designated by $d\sigma_a/dE$. 
Integrating the expression of Eq.~(\ref{brem}) over the BS05 standard solar
model \cite{Bach05} we find that the expected solar axion flux at the
Earth is $\Phi_a = g_{ae}^2\, 1.3\times 10^{35}\: {\rm cm}^{-2}\, 
{\rm s}^{-1}$ with an 
average energy of 1.6 keV and an approximate spectrum
\begin{equation} 
   \frac{d\Phi_a}{dE} =
   g_{ae}^2\,
  1.55 \times 10^{35}\,({E}_{\rm keV})^{0.63}\,
   e^{-E_{\rm keV}}\: 
   {\rm cm}^{-2}\,{\rm s}^{-1}\,{\rm keV}^{-1}\, , 
\label{aflux}
\end{equation}
as shown in Fig.~\ref{flux}. 
Here $E_{\rm keV}\equiv E/{\rm keV}$.
%%%%%%%%%%%%%%%%%%%%%%%
The corresponding solar axion luminosity is calculated to be
$L_a = g_{ae}^2\, 2.4\times 10^{20} L_{\odot}$, where $L_{\odot} = 
3.84\times 10^{26}\: {\rm W}$ is the solar photon luminosity.
%%%%%%%%%%%%%%%%%%%%%%%
In particular, the coupling exploited in our 
experiment, of the hadronic axion to electrons, is given numerically by
\cite{Lju04}
\begin{equation} 
  g_{ae}=6.6\times 10^{-15}\, m_{\rm eV}\left[\frac{E}{N}(23.2-
        {\rm ln}m_{\rm eV})-14.8 \right ] \, ,
\label{coupling}
\end{equation}
where $E/N$ is the model-dependent ratio of electromagnetic and 
color anomalies, 
and $m_{\rm eV} \equiv m_a/{\rm eV}$. For hadronic axions that have greatly 
suppressed photon couplings ($E/N \approx 2$) \cite{Kap85} this is
$g_{ae}\approx (2.0\rightarrow1.2)\times 10^{-13}\,m_{\rm eV}$ over a 
broad range of axion masses of $1\rightarrow1000$~eV.
%%%%%%%%%%%%%%%%%%%%%%%%%%%%%%%%%%%%%%%%%%%%%%%%%%%%%%%%%%%%%%%%%%%%
We note that astrophysical arguments related to stellar 
evolution ($L_a \lesssim L_{\odot}$)~\cite{Raf96} and helioseismology 
($L_a \lesssim
 0.2\,L_{\odot}$)~\cite{Sch99} would imply upper bounds on the 
axion-electron coupling of $6.5\times 10^{-11}$ and $2.9\times 10^{-11}$,
respectively. By using Eq.~(\ref{coupling}) they translate to
upper bounds on the hadronic axion mass of 515~eV and 210~eV. 
%%%%%%%%%%%%%%%%%%%%%%%%%%%%%%%%%%%%%%%%%%%%%%%%%%%%%%%%%%%%%%%%%%%%
%%%%%%%%%%%%%%%%%%%%%%%%%%%%%%%%%%%%%%%%%%%%%%%%%%%%%%%%%%%%%%%

\begin{figure}[htb!]
\centerline{\includegraphics[width=95mm,angle=0]{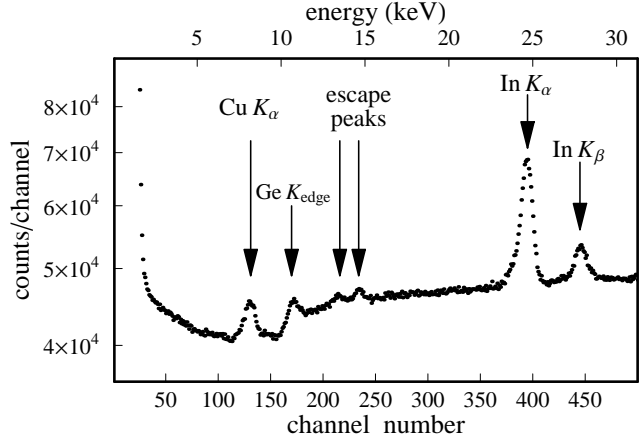}}
               \caption{Total measured energy spectrum
showing also x-ray peaks from various materials.}
  \label{data}
   \end{figure}

In this Letter, we report results of our search for 
solar hadronic axions which could be produced in the Sun by the 
bremsstrahlung-like process and detected in the single spectrum of an 
HPGe detector as the result of the axioelectric effect on
germanium atoms\footnote{Using theoretical predictions of \cite{Zhi79} we 
  estimate that for $\sim$keV axions the axioelectric effect is about three 
  orders of magnitude stronger than the axion-to-photon Compton process.}.
The x rays accompanying the axioelectric effect will be subsequently absorbed
in the same crystal, and the energy of the particular outgoing signal equals 
the total energy of the incoming axion. Because in this experimental set-up 
the target and detector are the same, the efficiency 
$\varepsilon$ of the system is 
substantially increased, i.e., $\varepsilon \approx 1$.  
The HPGe detector with an active target mass of 1.5~kg was placed 
at ground level, inside a 
low-radioactivity iron box with a wall thickness ranging from 16 to 23 cm. The
box was lined outside with 1 cm thick lead. The crystal was installed in a 
standard PopTop detector capsule (Ortec, model CFG/PH4) with a beryllium
window of 0.5 mm. 
The HPGe preamplifier signals were distributed to a spectroscopy amplifier
at the 6~$\mu$s shaping time.
A low threshold on the output provided the online trigger, ensuring that
all the events down to the electronic noise were recorded.
The linearity and energy resolution have been studied by using various 
calibrated sources and, in particular, in the lowest-energy 
region mainly a $^{241}$Am source
(13.9 keV x-rays and their escape peak of 3.9~keV). 
Data were accumulated in a 1024-channel analyzer, with an energy dispersion of
63.4~eV/channel, in 20-hour cycles 
with total time of collection of 275 days. 
In long-term running conditions, the
knowledge of the energy scale is assured by continuously monitoring
the positions and resolution of the In x-ray peaks which are present
in the measured energy distribution. Drifts were 
$<\pm 1$ channel and a statistical accuracy of better than 0.6\% per 
channel was attained. Thus, all of the spectra were summed together without
applying any gain shifts or offsets and are shown in Fig.~\ref{data}.
%%%%%%%%%%%%%%%%%%%%%%%%%%%%%%%%%%%%%%%%%%%%%%%%%%%%%%%%%%%%%%%
%%%%%%%%%%%%%%%%%%%%%%%%%%%%%%%%%%%%%%%%%%%%%%%%%%%%%%%%%%%%%%%
Energy resolution (FWHM) was estimated to have a value of 
660~eV at the photon
energy in the 2.0~-~3.8 keV region, where the axion signal is expected. 
The former bound is imposed as the analysis threshold due to the
electronic noise while the latter one is due to the energy distribution of
the solar axions. 

Our experiment involves searching for the particular energy spectrum
in the measured data,
\begin{equation}
  S_n=2\, N_{\rm Ge}\,t \int_{E_n}^{E_{n+1}}dE^{'}
       \int_0^{\infty}\frac{d\Phi_a}{dE}\sigma_{ae\rightarrow e}
               \,R(E^{'},E)dE \:,
\label{signal}
\end{equation}
produced if the solar axions are detected via axioelectric effect.
Here $S_n$ is the number of counts detected in detector energy channel $n$,
$N_{\rm Ge}$=1.24$\times 10^{25}$ is the number of germanium atoms in our
67~mm in diameter and 80~mm thick HPGe crystal, and $t$ is 
the time of measurement. Factor 2 is number of
electrons in (L$_1$-) M$_1$-shell while their binding energies are 
$E_{\rm b}$(L$_1$)=1.413 keV and $E_{\rm b}$(M$_1$)=0.181 keV.
The axion response function of the detector
$R(E^{'},E)$ is well represented by a Gaussian (to describe full energy peak) 
with the amplitude, normalized
on the efficiency $\varepsilon$, and width of 0.42\,FWHM. 
Using Eq.~(3) from \cite{Lju04}, the cross section 
for axioelectric effect per electron
was calculated including contributions  from 2$s$ and 3$s$ initial state
electrons and may be written in the $m_a\rightarrow 0$ \& $E\ll m_e$ 
approximation and for $E>E_{\rm b}$(L$_1$) as: 
\begin{equation}
 \sigma_{ae\rightarrow e}=\left(\frac{1}{2^3}+\frac{1}{3^3}\right)
   \frac{\sqrt{2}}{4\,\pi}\,{g_{ae}^2}\,\sigma_{\rm Th}\,{\alpha}^3
     Z^5\, \left(\frac{m_e}{E}\right)^{3/2}\:.
\label{cross}
\end{equation}
Here $\sigma_{\rm Th}=6.65\times 10^{-25}$ cm$^2$ is the cross 
section for Thomson
scattering, $\alpha$=1/137 is the fine-structure constant, $Z$=32 
is the atomic number of Ge, and $m_e$ is the electron mass.  
Equation (\ref{cross}) is consistent with a more general expression
$\sigma_{ae\rightarrow e}=(\alpha_{ae}/\alpha)\, (E/2m_e)^2\,
\sigma_{\rm ph}$ \cite{Dim86}, where $\alpha_{ae}=g_{ae}^2/4\,\pi$
and $\sigma_{\rm ph}$ is the photoelectric cross section.

We have used the fit method to analyse the data in order to find
upper limits on $g_{ae}$ ($m_a$), as described below.   
Similar approaches have been used by other groups~\cite{Bau98} 
to extract (conservative) upper limits
in a case like this where direct background measurement is not possible
(the Sun cannot be switched off) and the signal shape is a broad smooth
spectrum on top of an unknown background spectrum.

\begin{figure}[htb!]
\centerline{\includegraphics[width=95mm,angle=0]{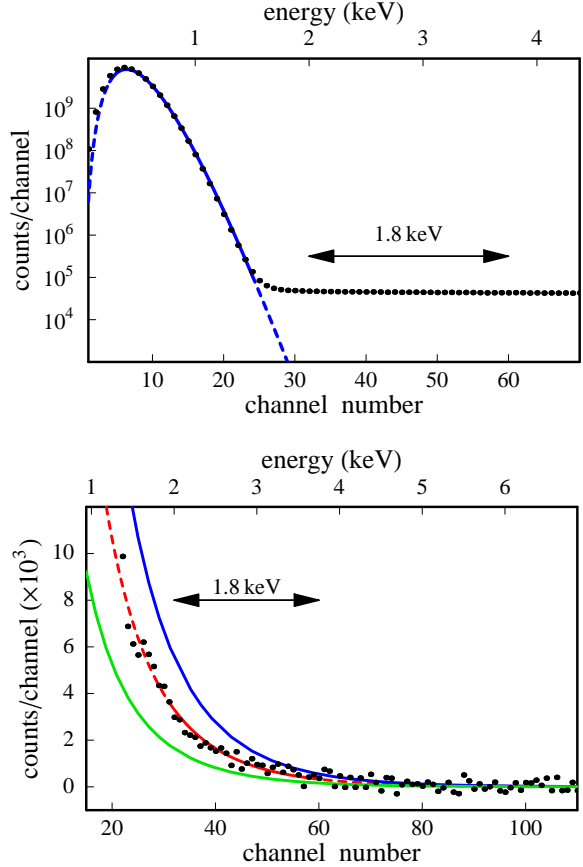}}
               \caption{
Top panel: low-energy data are shown together with the
best fit for the electronic noise expectation 
$T_n$ (solid line) and its 
extrapolations (dashed line). 
Bottom panel: residuals from the noise and background expectation are
shown together with the expectations for best fit $g_{ae}=4.4\times 10^{-11}$
(red line) as well as for $g_{ae}=5.0\times 10^{-11}$ (blue line) and
$g_{ae}=3.7\times 10^{-11}$ (green line). 
On both panels arrows indicate the energy region of interest for axion signal.} 
  \label{fit}
   \end{figure}

In the fit method we have used the following 
procedure to make the estimation of an axion signal.
%%%%%%%%%%%%%%%%%%%%%%%%%%%%%%%%%%%%%%%%%%%%%%%%%%%%%%%%%%%%%% 
The experimental data in the energy interval
2.0~-~3.8 keV (corresponded to MCA channels of 32 to 60) 
was fitted by the sum of three functions:   
the electronic noise expectation ($T_{n}$), the background expectation 
($B_n$) and the
effect being searched for ($S_n$). To estimate the electronic noise
the data were fit to a four-parameter function in 
the region below 2 keV 
(corresponded to the channels of 4 to 24); it has the form
$T_n=A\,n^B{\rm exp}(Cn^D)$, 
where $A=1.46\times 10^7$, $B=6.85$, $C=-0.858$, and $D=1.08$, 
see Fig.~\ref{fit}~(top panel).  
As a simple background estimation, suitable for the present purposes, 
a first-order polynomial $B_n=a+bn$ has been assumed.
Values of $a=4.54\times 10^4$ and 
$b=-42.5$ were found by fitting $B_n$ with
the data in the region above 3.8 keV, corresponded to the  
channels of 70 to 110. From the current limits on the axion mass
\cite{Nam07,Der07} we estimate that the axion signal is
negligible in this channel interval. 
Contribution of axions was then investigated
by extrapolating the electronic noise $T_n$ and background $B_n$ 
in the 2.0~-~3.8 keV region, where the detection of
axion events should be the most efficient. Inserting Eqs.~(\ref{aflux}) and 
(\ref{cross}) in Eq.~(\ref{signal}), one can express the expected 
axion spectrum $S_n$ as a function of $g_{ae}^4$.
As the best values of 
$A$, $B$, $C$, $D$, $a$ and $b$
had been already determined in the regions where
axion contributions could be neglected, only $g_{ae}^4$ was varied 
in the ${\chi}^2$ comparison. 
We were using $g_{ae}^4$ instead of $g_{ae}$ as the minimization parameter
because the signal strength (i.e., number of counts) is proportional to
$g_{ae}^4$. The results of the analysis are consistent
with $g_{ae}^4= (3.6 \pm 0.1)\times 10^{-42}$ at 1$\sigma$ level.
Figure~\ref{fit}~(bottom panel) displays the results of our fit.
%%%%%%%%%%%%%%%%%%%%%%%%%%%%%%%%%%%%%%%%%%%%%%%%%%%%%%%%%%%%%%%
%%%%%%%%%%%%%%%%%%%%%%%%%%%%%%%%%%%%%%%%%%%%%%%%%%%%%%%%%%%%%%%
Since we cannot exclude more complicated scenarios in which both the
background and  the electronic noise modify 
significantly the energy spectrum of signal events in the region of 32 
to 60 channels, we treat the   
residuals from the noise and background expectation 
conservatively as an upper limit on the axion signal.
In other words, we adopted a criteria commonly used in experiments where direct background measurement is not possible \cite{Bau98} that the theoretically 
expected signal cannot be larger than that observed at a given confidence 
level.
In this way we obtain 
the standard 95\% C.L. upper limit \cite{Yao08} 
$g_{ae} \lesssim 4.4\times 10^{-11}$
which translates through Eq.~(\ref{coupling}) to $m_a \lesssim 334$~eV.
Because the precise form of the background expectation is not known, 
we tried also other plausible forms, such as
$B_n=a+bn+cn^2$ and $B_n=a+bn+cn^{-1}$.
The former yields 
$g_{ae}^4=(2.5\pm 0.2)\times 10^{-42}$ resulting in $m_a \lesssim 310$~eV 
(95\% C.L.) while the latter provides  
$g_{ae}^4=(1.4\pm 1.9)\times 10^{-43}$ implying the more stringent 
limit $m_a \lesssim 190$~eV (95\% C.L.).  
One can see that subtraction of
the background expectation in a form of $a+bn+cn^{-1}$ 
cancels the excess of events and the obtained result is very close to the
sensitivity of the experiment. 
Namely, if we assume that the measured spectrum in the region
of interest for axion signal is compatible with the background
expectation, the fluctuation due to statistical uncertainty
imposes a limit on the maximum allowable number of axion events. 
In this case (the mean is zero and statistical uncertainty for 
$g_{ae}^4$ is $\sigma=2\times 10^{-43}$) 
we obtain $g_{ae} \lesssim 2.4\times 10^{-11}$ 
corresponding to $m_a \lesssim 170$~eV at the 95\% confidence level.

In conclusion, our measurements based on the coupling of the hadronic axions
to electrons set a conservative upper limit on the axion mass of
$m_a \lesssim 334$~eV at 95\% confidence level.
This upper limit is comparable to the current limits \cite{Nam07,Der07}  
which are extracted from the data taken 
in the experiments based on the axion-nucleon coupling. We point out 
the derived limit is free from the large uncertainties and ambiguity, 
associated with the
estimation of the flavor-singlet axial-vector matrix element, which are 
related to experiments based on the axion-nucleon coupling. 
New experiments
with lower electronic noise and reduced background could improve our
search for the hadronic axions.
Note that the axioelectric absorption peak in germanium occurs at lower
incoming axion momentum when $E$ coincides with the binding energy of 
atomic shells. At these energies, atomic bound state effects lead to large
enhancements in the detection rates of axions, similarly to those in
the photoelectric effect. For example,
the expected number of axion events for $E>1.4$~keV is about 3 times that
for $E>2$~keV. Thus an improvement in the discovery potential may come from
examination of the Ge data for energies near the L$_1$-shell peak of 1.413 keV.
Upgradings of the set-up to reduce the electronic noise and permit lowering
of the energy threshold below 1 keV are already foreseen and in preparation. In
the near future we expect that measurements with the 
array of 10~-~20 ultra-low-energy IGLET germanium detectors with
an anti-Compton shield could
improve significantly our understanding on hadronic axions.

We acknowledge support from the Ministry of Science, Education and Sports of
Croatia under grant no. 098-0982887-2872.
%%%%%%%%%%%%%%%%%%%%%%%%%%%%%%%%%%%%%%%%%%%%%%%%%%%%%%%%%%%%%%%%%%%%%%%%

\end{document}